\documentstyle[iopconf1,psfig]{article}

\begin{document}
\hspace*{\fill} LMU--TPW--98/13, MPI--PhT--98/61\\
\hspace*{\fill} hep-ph/9808245\\[3ex]

\title{Superstrings and dark matter}

\author{R Dick\dag, N Eschrich\dag\ and M Gaul\dag\ddag}

\affil{\dag\ Sektion Physik der Ludwig--Maximilians--Universit\"at,\\
Theresienstr. 37, 80333 M\"unchen, Germany}

\affil{\ddag\  Max--Planck--Institut f\"ur Physik, \\
F\"ohringer Ring 6, 80805 M\"unchen, Germany}

%%%%%%%%%%%%%%%%%%%%%%%%%%%%%%%%%%%%%%%%%%%%%%%%%%
% You may repeat \author \address as often as necessary      %
%%%%%%%%%%%%%%%%%%%%%%%%%%%%%%%%%%%%%%%%%%%%%%%%%%

\beginabstract
We point out that the spectrum and interactions of light states
of the heterotic string indicate a string scale close to the GUT scale, 
and a mass
generating scale for the gravitationally interacting states around $10^9$GeV
if these states
contribute a large fraction to dark matter.
\endabstract

%\section{Introduction}

\footnotetext{Invited talk, 2nd International Conference on Dark Matter
 in Astro and Particle Physics,
 Heidelberg (Germany) 20--24 July 1998.}

The apparent discrepancy between the amount of energy that may exist in
the form of baryons
and the amount of energy that is needed
for structure formation and to explain observations of gravitational
 lensing and
peculiar flows of large scale structures 
constitutes one of the most exciting scientific puzzles
at the turn of the century.  
The clarification of the amount and composition of the dark matter
in the universe calls for the joint efforts
of astronomers, astrophysicists, relativists and particle physicists,
and has implications for our understanding of physics both
at the largest and the smallest scales.

The success of standard Big Bang nucleosynthesis
and a lower bound on the Hubble
parameter $H_0=
100h\frac{\mbox{\small km}}{\mbox{\small s}\cdot\mbox{\small Mpc}}
 \ge 50\frac{\mbox{\small km}}{\mbox{\small s}\cdot\mbox{\small Mpc}}$ 
together with observational constraints on Helium and Deuterium abundances
put a strong
constraint $\Omega_B\leq 0.024h^{-2}<0.1$ on the baryonic energy
 density \cite{ST}, and the survey of baryonic matter 
by Fukugita, Hogan and Peebles \cite{FHP} implies 
 $\Omega_B\leq 0.0276h^{-1}+0.0093h^{-1.5}\leq 0.082$.
On the other hand, estimates from galaxy clusters \cite{cnoc}, 
intracluster gas fractions \cite{WE},
 strong gravitational lensing \cite{maer}
and peculiar velocities
of galaxies \cite{dekel} 
all indicate  that there must be more energy in matter than
can be stored in baryons: A 
conservative lower bound is $\Omega_M\ge 0.15$. 
Numerical investigations of structure evolution caution us that
galaxy clusters 
may not directly trace the mass distribution and that observed peculiar
velocity fields may agree with cosmological models for a wide
range of matter energy densities \cite{sw}, but they also show
that non-baryonic cold dark matter 
is an indispensable ingredient in forming the observed structure
in the universe: Pressureless matter must have dominated
the energy density of the universe well before baryon--radiation
equality for the density contrast to have evolved into the non-linear
regime. For a
critical and thorough recent survey
of the evidence for dark matter see \cite{gb}.

Particle physics scenarios
for non-baryonic dark matter can roughly be classified into
bottom--up or  top--down approaches,
 starting either
from minimal extensions of the standard model or
 from promising {\it Ans\"atze}
for particle physics at the Planck scale. Supersymmetric
 extensions of the standard model
containing a lightest supersymmetric particle by $R$--parity  
or inclusion of an anomalous 
 $\mbox{U}(1)$--symmetry implying existence of a weakly coupled
 pseudoscalar axion 
provide interesting examples for the bottom--up approach.

 For top--down approaches to dark matter and physics beyond the standard model
the heterotic string of Gross, Harvey, Martinec and Rohm still provides
 an interesting
starting point because it 
makes definite predictions about the spectrum of excitations and symmetries
below but close to the
quantum gravity scale \cite{GHMR,GSW} . 
Even in the framework of M--theory the heterotic string provides
an inevitable step towards low energy phenomenology \cite{hw},
 and if we are willing to
accept the extrapolation of supersymmetric $\beta$ functions over thirteen
orders of magnitude on the energy scale
weakly coupled heterotic string theory still provides
a compelling scenario for GUT scale physics including gravity.

Another interesting approach employs minimal SUGRA
unified models \cite{AN}. These models can be motivated independently
from string theory, but they are also clearly relevant for 
supersymmetric dark matter
in the visible sector of the heterotic string, and in
particular for the problem whether there is an LSP
contribution.
Besides an LSP,
superheavy particles 
may also contribute
if mass bounds are avoided through
non-thermal creation at the end of inflation \cite{heavy}.
Benakli, Ellis and Nanopoulos implemented this mechanism in a string model
where it yields superheavy bound states in the hidden sector \cite{BEN}.

In the sequel we provide estimates on two scales that arise 
in heterotic string theory due to the large number 
and interactions of light helicity states: The fact that the majority of
states interacts strongly enough to be thermalized already at high
scales indicates that a radiation dominated heat bath
emerges at the string scale. As a consequence, this scale turns
out to be close to the GUT scale, well below the Planck scale.
  Furthermore, apart from the graviton those states 
which interact only with gravitational strength have to
acquire mass terms. One can give an upper bound
on the temperature where these mass terms arise from 
the requirement that
the universe is open or flat.

In explaining these points
we will rely on the following assumptions:
All states in the theory arise from excitations of fundamental closed
 strings, and
in describing physics near and below the GUT scale we may neglect the
 massive string excitations
which are separated by a mass gap of order
 $m_{Pl}=(8\pi G_N)^{-1/2}=2.4\times 10^{18}$GeV. Furthermore, 
no Kaluza--Klein scale 
is taken into account: Space-time is assumed to
be four-dimensional below the string scale.

In the original formulation internal symmetries are constrained 
to gauge groups SO$(32)$ 
or $\mbox{E}_8\times\mbox{E}_8$, and in the sequel the phenomenologically
more interesting $\mbox{E}_8\times\mbox{E}_8$
theory is preferred.
However, a qualification to be kept in mind concerns the possibility
to change the world sheet degrees of freedom in a way which maintains
mathematical consistency of the theory, see \cite{tye} and references there.
This can enlarge 
the  rank of the gauge group and
change the ratio of states with or without gauge interactions.

In the $\mbox{E}_8\times\mbox{E}_8$ theory
the spectrum of massless or light degrees of freedom
at high energies  
comprises 8064 helicity states\footnote{Each fermionic
helicity state is counted twice corresponding to two real on-shell degrees
of freedom per helicity state.}
with the following
multiplet structure:\\[0.5ex]

\begin{tabular}{ll}
$(1,1)$ & 1 graviton, 1 gravitino, 1 axion, 1 dilaton, 1 axino/dilatino,\\
{}& 12 vectors, 36 scalars, 30 Weyl fermions.\\
$(248,1)$ & 248 $\mbox{E}_8$--gluons and 248 $\mbox{E}_8$--gluinos
 in one multiplet,\\
{}&  744 complex scalars
  and 744 Weyl fermions in 3 multiplets.\\
$(1,248)$ & as above.\\
\end{tabular}
\vspace{1.5ex}

Depending on the starting point for the formulation of a four--dimensional
effective action the determinant of 21 of the 36 
real scalars in the $(1,1)$ sector besides
the string dilaton couples like a Kaluza--Klein dilaton, and a particular
combination of
the remaining 15 isoscalar scalars
couples like a further axion\footnote{In the 
approach to four--dimensional supersymmetric low energy models
through Calabi--Yau manifolds
the symmetry between the two $\mbox{E}_8$ sectors is broken by hand
by embedding an SU$(3)$ spin connection in one $\mbox{E}_8$, thus breaking
the gauge group to $\mbox{E}_6\times\mbox{E}_8$ and eliminating
 32 helicity states.}.

Superficially,
the number of 8064 helicity states of the $\mbox{E}_8\times\mbox{E}_8$ 
heterotic string
seems very large compared to the 120--126 helicity states of the
standard model including gravity, and occasionally this is referred
to as an ``embarrassment of riches''. However, 
if one thinks about it more thoroughly the number of light states
predicted by the heterotic string is surprisingly small: 
It implies that ``on average" we will have to increase the energy
by more than $10^9$GeV to encounter one new
degree of freedom!

The spectrum and gauge symmetries of the light states allow
us to infer
cosmologically relevant features of the heterotic string
independent from the details of the four--dimensional
effective action: A gauge coupling of order $\alpha_G\simeq 0.04$
at the GUT scale implies that all the states in adjoint $\mbox{E}_8$
multiplets are coupled strongly enough to constitute
a thermalized heat bath, and since these states originate as massless
states in string theory it is safe to say that this heat bath
is radiation dominated. The heat bath in fact corresponds to two
components interacting only weakly with gravitational strength.
However, due to the symmetry of the two $\mbox{E}_8$ sectors
it seems reasonable to expect that temperature differences
between both sectors can be neglected.
Immersed in the heat bath there are 128 helicity states
with only gravitational strength couplings, and these are certainly too
weakly coupled to be thermalized, yet they come massless. Therefore, we expect
that below but close to the scale where the finite extension of strings
must be taken into account the energy density can be described
by a mixture of a stiff fluid with dispersion relation $p_\phi=\rho_\phi$
and dominating radiation with pressure $p_\gamma=\rho_\gamma/3$.
Due to its large pressure a stiff fluid does a lot of work during the expansion
of the universe and its energy density drops with the scale parameter $R$
according to $\rho_\phi\sim R^{-6}$.

In a mixture of radiation and a stiff fluid 
the comoving time $t$ is related to the scale parameter 
according to 
\[
\frac{2}{\sqrt{3}m_{Pl}}\rho_{0\gamma}(t-t_0)=
x\sqrt{x^2\rho_{0\gamma}+\rho_{0\phi}}
-\sqrt{\rho_{0\gamma}+\rho_{0\phi}}
\]
\[
-
\frac{\rho_{0\phi}}{\sqrt{\rho_{0\gamma}}}
\ln\!\Big(
\frac{x\sqrt{\rho_{0\gamma}}+\sqrt{x^2\rho_{0\gamma}+\rho_{0\phi}}}
{\sqrt{\rho_{0\gamma}}+\sqrt{\rho_{0\gamma}+\rho_{0\phi}}}
\Big), 
\]
where $\rho_{0\gamma}$ and $\rho_{0\phi}$ denote the energy densities
in radiation and decoupled massless states at the fiducial time $t_0$,
and $x=R/R_0=R(t)/R(t_0)$.

However, in order not to generate new long range forces at the present epoch
the massless helicity states in the gravitational sector besides the graviton
must have acquired mass terms, and then the energy
density in these states
decays slower than the surrounding heat bath: $\rho_\phi\sim R^{-3}$.
This behaviour is independent from thermalization
and implies that the generation of mass terms could not happen too
early without contradicting the widely accepted upper 
bound $\rho\le\rho_{k=0}\simeq 81h^2\mbox{meV}^4$ on the present energy density.
On the other hand, the very weak coupling and the slow decay
of these states 
make them ideal candidates for cold dark matter in the universe, and
we can estimate the transition temperature $T_c$ under 
the following provisos \cite{DG}:\\
-- The massive states emerging from the gravitational sector
of the heterotic string generate a large CDM contribution 
to the present energy density.\\
-- Discontinuities in $\rho_\phi$ during the transition 
can be neglected.\\
-- Radiative modes which become massive decay efficiently enough
to keep the product $g_\ast T^4$ approximately continuous.
This concerns most of the modes, many of which must acquire
mass terms already close to the string scale\footnote{If all but the helicity states
of the minimal supersymmetric standard model
would acquire masses at a single scale,
and if the massive modes would release their energy
instantaneously through decay
or conversion into the remaining light
degrees of freedom, this would result in a temperature increase 
by a factor 2.}.\\[-1.5cm]
\begin{figure}[ht]
\centerline{\psfig{figure=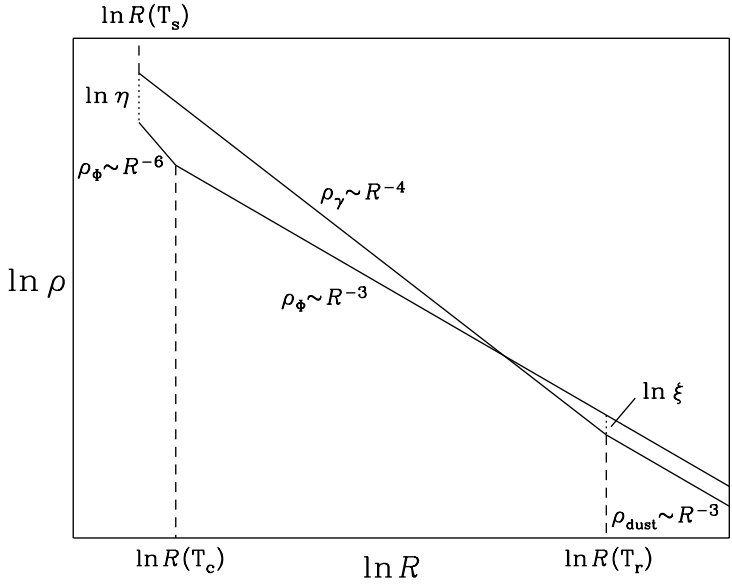,width=10cm}}
\caption{Transition from a stiff fluid to CDM. \label{fig1}}
\end{figure}

Denoting by  $\eta=\rho_{\gamma}(T_s)/\rho_{\phi}(T_s)$ the ratio
of energy densities at the string scale $T_s$
and by $\xi=\rho_{\phi}(T_r)/\rho_{\gamma}(T_r)$ the ratio of energy densities
at baryon--radiation equality we find:
\begin {equation}\label{tc}
T_c=(\eta\xi T_r T_s^2)^{1/3}.
\end{equation}

Before employing this formula to estimate $T_c$ we stress that the large
number of thermalized helicity states 
corresponding to $g_\ast=7440$ implies a string scale
 $T_s$ in coincidence with estimates of the GUT scale from extrapolations
of supersymmetric $\beta$ functions: In the usual approximation of
an ideal gas the energy density in the thermalized states is
\[
\rho_\gamma=\frac{3m_{Pl}^2}{4t^2}=\frac{\,\pi^2}{30}g_{\ast}T^4.
\]
If we now require that the typical wavelength does 
not exceed the horizon or the age of the
universe we find a maximal temperature
\[
T\le T_s=
\sqrt{\frac{45}{2g_\ast}}\frac{m_{Pl}}{\pi}\simeq 4\times 10^{16}\mbox{GeV}.
\]
Beyond this temperature a particle 
description makes no sense and a reasonable guess
is to identify this temperature with the scale 
where a string description must take over.
However, we will consider 
the range $4\times 10^{16}\mbox{GeV}\le T_s\le m_{Pl}
=2.4\times 10^{18}\mbox{GeV}$
in (\ref{tc}). For $\eta\simeq 60$ 
(equipartition at the string scale), $\xi\simeq 10$ 
and $T_r\simeq 0.3$eV
we then find $7\times 10^{8}\mbox{GeV}\le T_c\le 1\times 10^{10}\mbox{GeV}$.
It is also of interest to estimate the mass scale emerging at $T_c$: 
Since the typical coupling scale
for the gravitational states is $m_{Pl}$ we expect masses
\[
m\simeq\frac{T_c{}^2}{m_{Pl}}
\]
in the range between 200MeV and 40GeV.

The emergence
of massive states of only gravitational coupling strength
is a generic possibility in string theory, and this is a matter of concern
for dark matter searches: We cannot exclude the possibility
that a considerable fraction of dark matter couples to ordinary matter
so weakly that we may notice it only through large scale gravitational
effects, but not in dedicated particle physics experiments.
\vspace{-7pt}
\section*{Acknowledgements}
RD and NE acknowledge support by the DFG
through SFB 375--95 and GK 7--93, respectively.
RD would also like to thank the dark matter group in Heidelberg
for the invitation and for hospitality during a very interesting meeting.


\vspace{-7pt}
\newpage\noindent
\begin{thebibliography}{88}
 \bibitem{ST}Schramm D N and Turner M S 1998
   {\it Rev.\ Mod.\ Phys.\ }{\bf 70} 303
 \bibitem{FHP}Fukugita M, Hogan C J and Peebles P J E 1997
   {\it The cosmic baryon budget} astro--ph/9712020
 \bibitem{cnoc}Carlberg R G, Yee H K C and Ellingson E 1997
  {\it Astrophys.\ J.\ }{\bf 478} 462
  \newref Carlberg R G et al 1998
  {\it The $\Omega_M$ -- $\Omega_\Lambda$ constraint from CNOC clusters}
    astro--ph/9804312
 \bibitem{WE}White S D M, Navarro J F, Evrard A E and Frenk C S 1993
    {\it Nature }{\bf 366} 429
   \newref Evrard A E 1997 {\it MNRAS }{\bf 292} 289
 \bibitem{maer}Mellier Y, Bernardeau F and van Waerbeke L 1998
  {\it Dark matter and gravitational lensing} astro--ph/9802005
 \bibitem{dekel}Sigad Y, Eldar A, Dekel A, Strauss M A and Yahil A 1998
  {\it  Astrophys.\ J.\ }{\bf 495} 516
 \bibitem{sw}Jenkins A, Frenk C S, Pearce F R, Thomas P A , Colberg J M, White
   S D M, Couchman H M P, Peacock J A, Efstathiou G and Nelson A H
   1998 {\it Astrophys.\ J.\ }{\bf 499} 20
   \newref Thomas P A, Colberg J M, Couchman H M P, 
    Efstathiou G P, Frenk C S, 
      Jenkins A R, Nelson A H, Hutchings R M, Peacock J A, 
     Pearce F R and White S D M 1998 {\it MNRAS }{\bf 296} 1061
 \bibitem{gb}B\"orner G 1997
 {\it Beyond the Desert -- Accelerator and Non-Accelerator Approaches}
  (Bristol: IOP Publishing) p~769
 \bibitem{GHMR}Gross D J, Harvey J A, Martinec E and Rohm R 1985 
 {\it Nucl. Phys. }{\bf B256} 253, 1986 {\it Nucl. Phys. }{\bf B267} 75
 \bibitem{GSW}Green M B, Schwarz J H and Witten E 1987
  {\it Superstring Theory} 2 Vols (Cambridge University Press)
 \bibitem{hw}Ho\v{r}ava P and Witten E 1996 {\it Nucl.\ Phys.\ }{\bf B460} 506,
 {\it Nucl.\ Phys.\ }{\bf B475} 94
 \bibitem{AN}Pran Nath and Arnowitt R 1997 {\it Phys.\ Rev.\ }{\bf D56} 2820
 \bibitem{heavy}Chung D J H, Kolb E W and Riotto A 1998 
  {\it Superheavy dark matter} hep--ph/9802238,
  {\it Nonthermal supermassive dark matter} hep--ph/9805473
 \newref Kuzmin V A and Tkachev I I 1998 {\it Ultra-high energy cosmic rays, 
    superheavy long-living particles, and matter creation after inflation}
    hep--ph/9802304
 \bibitem{BEN}Benakli K, Ellis J and Nanopoulos D V 1998
  {\it Natural candidates for superheavy dark matter in string and M theory}
  hep--ph/9803333
 \bibitem{tye}Kakushadze Z, Shiu G, Tye S-H H and Vtorov--Karevsky Y 1998
  {\it Int.\ J.\ Mod.\ Phys.\ }{\bf A13} 2551 
 \bibitem{DG}Dick R and Gaul M 1998
  {\it Cosmological implications of a light dilaton} hep--ph/9801249
\end{thebibliography}
\end{document}